\documentclass[fleqn,twoside]{article}
\usepackage{espcrc2}

\usepackage{epsfig}

\newcommand{\AmS}{{\protect\the\textfont2
  A\kern-.1667em\lower.5ex\hbox{M}\kern-.125emS}}

\newcommand{\bi}{\scriptstyle b_{1}}

\newcommand{\hi}{\scriptscriptstyle h_{1}}

\newcommand{\no}{\nonumber}

\newcommand{\delv}{\delta q^v}

\newcommand{\ai}{\scriptstyle a_{1}}

\newcommand{\bgi}{\begin{itemize}}
\newcommand{\eni}{\end{itemize}}

\newcommand{\be}{\begin{eqnarray}}
\newcommand{\ba}{\begin{array}}
\newcommand{\ea}{\end{array}}
\newcommand{\ee}{\end{eqnarray}}

\newfont{\fib}{cmfi10 at 10pt}

\newcommand{\eg}{{\it e.g.}\ }

\title{Transversity and Meson Photoproduction}

\author{Gary R. Goldstein\address{Department of Physics 
          and Astronomy, Tufts University, 
           Medford, MA 02155, USA} 
        \thanks{Research suported in part by a grant from the US 
Department of Energy DE-FG02-29ER40702.}
        Leonard Gamberg\address{Physics Department,
Penn State University, Berks-Lehigh Valley College, Reading, PA 19610, 
USA}}

\begin{document}

\begin{abstract}
Both meson photoproduction and semi-inclusive deep
inelastic scattering can potentially probe transversity
in the nucleon.  We explore how that potential can be
realized dynamically.  The role of rescattering in
both exclusive and inclusive meson production as a source
for transverse polarization asymmetry is examined. We use
a dynamical model to calculate the asymmetry and relate
that to the transversity distribution of the nucleon.
\vspace{1pc}
\end{abstract}

\maketitle


The leading twist transversity distribution $h_1(x)$~\cite{jaffe91} and 
its first moment, the tensor charge, are as fundamental to our 
understanding of the spin structure of the
nucleon as are the helicity distribution and the axial 
vector charge.  Unlike $g_1(x)$, though, the chiral
odd  $h_1(x)$ cannot be accessed in deep inelastic scattering.  
However, $h_1(x)$ can be probed when  at least two hadrons are 
present, \eg Drell Yan~\cite{ralston79} or
semi-inclusive deep inelastic scattering (SIDIS). In the
latter process at leading twist, the effect of quark transversity
can be measured via the azimuthal asymmetry
in the fragmenting hadron's momentum and spin distributions.
For spinless hadrons 
the so-called Collins asymmetry~\cite{cnpb92_1} 
 depends on the transverse momentum of quark
distributions in the target and fragmentation
functions for outgoing hadrons~\cite{kotz0}. 
Including transverse momentum 
leads to an increase in the number of
leading twist distribution and fragmentation functions and can 
involve $T$-odd quark 
functions~\cite{mulders2}.

Non-zero transverse single spin asymmetries (SSA) have been measured at 
HERMES and SMC in semi-inclusive pion electroproduction~\cite{hermes}.   
These data could point to the essential role played by quark
transverse
momenta and $T$-odd distributions. Recently, further insight into  
transversity has come from the interpretation of deeply virtual Compton 
scattering (DVCS)~\cite{diehl}  where the
quark target helicity flip amplitudes,
written in terms of the
generalized parton distributions (GPD) $H^a_T(x,\xi,t)$,
reduce in the forward limit to the ordinary transversity
distribution, $\delta q^a(x)$.
Angular momentum conservation in these 
amplitudes requires that helicity changes are 
accompanied by a transfer of 1 or 2 
units of orbital angular momentum,
highlighting the essential role played 
by the $k_{\perp}$ generalizations of the quark transversity distribution.
The $t\rightarrow 0$ limit of
the associated form factor  is the tensor charge. 

This interdependence of transversity on
quark {\em orbital} angular momentum and $k_{\perp}$
is more general than the GPD analysis of 
transversity. This behavior arises
in ref.~\cite{gamb_gold} where we study the 
vertex function associated with the tensor charge. Again, angular 
momentum conservation results in  the transfer of 
orbital angular momentum $\ell=1$ carried by the dominant 
$J^{PC}=1^{+-}$
mesons  to compensate for the non-conservation of helicity across
the vertex. 
Transverse momentum dependence arises from
the axial vector mesons that dominate the
tensor coupling.~\footnote{$C$-odd mesons -- $h_1$(1170), $h_1$(1380), 
$b_1$(1235)}
These mesons are in the $\left(35\otimes \ell=1\right)$ 
multiplet of the 
$SU(6)\otimes O(3)$ symmetry group that best represents the
mass symmetry among the low lying mesons.
Along with axial vector
dominance this symmetry results in the isoscalar and isovector
contribution to the tensor charge 
\small\be
\delta u -\delta d
=\frac{5}{6}\frac{g_A}{g_V}\frac{ M_{\ai}^2}{
M_{\bi}^2}\frac{\langle k_{\perp}^2\rangle}{ M_N M_{\bi}}, 
\, \delta u+\delta d=\frac{3}{5}\frac{M_{\bi}^2}{M_{\hi}^2}\delv .
\nonumber
\ee\normalsize
Each depends on two powers of the average intrinsic quark
momentum $\langle k_{\perp}^2\rangle$, because  the tensor couplings
involve helicty flips that require kinematic factors
of $3$-momentum transfer.

The $k_{\perp}$ dependence can be understood on fairly general grounds
from the kinematics of the exchange picture
in exclusive pseudoscalar meson photoproduction. For large
$s$ and relatively small momentum transfer $t$ simple combinations of 
the 
four helicity amplitudes involve definite parity exchanges. 
The four independent helicity amplitudes can have
the minimum kinematically allowed powers,
\small\be
f_1 = f_{1+,0+} \propto k_{\perp}^1 , \
f_2 = f_{1+,0-} \propto k_{\perp}^0 , \nonumber 
\\
f_3 = f_{1-,0+} \propto k_{\perp}^2 , \
f_4 = f_{1-,0-} \propto k_{\perp}^1.  \nonumber
\ee
\normalsize
However, in single hadron exchange (or Regge pole exchange)
parity conservation requires
\small\be
f_1 =\pm f_4 \quad {\rm and} \quad f_2 =\mp f_3
\nonumber
\ee\normalsize
for even/odd parity exchanges. These pair relations,
along with a single hadron exchange model,
force $f_2$ to behave like  $f_3$ for small
$t$. This introduces the $k_{\perp}^2$ factor into $f_2$. However for a
non-zero polarized target asymmetry  to arise there must be
interference between single helicity flip and non-flip and/or double 
flip
amplitudes. Thus this asymmetry must arise from rescattering
corrections (or Regge cuts-eikonalization or loop corrections 
) to single hadron exchanges. That is, one of the amplitudes in
\small\be
P_y = \frac{2 Im (f_1^*f_3 - f_4^*f_2)}
{\sum_{j=1...4}|f_j|^2}
\nonumber
\ee\normalsize
must acquire a different phase.
Rescattering  reinstates $f_2 \propto k_{\perp}^0$ by integrating over loop
$k_{\perp}$, which
effectively introduces a $\left< k_{\perp}^2 \right>$ 
factor~\cite{gold_owens}.
This is true for the {\em inclusive process} as well,
where only one final hadron is measured; a relative phase
in a helicity flip three body amplitude is required.

Recently a rescattering approach was applied to the calculation of 
SSA in pion electroproduction, $e p\rightarrow e 
\pi X$,
using a QCD motivated quark-diquark model of the nucleon~\cite{brodsky} 
(BHS).
In Ref.~\cite{ji,cplb} the rescattering effect
is interpreted as giving rise to the $T$-odd Sivers $f_{1T}^\perp$ 
function;
 the number density of upolarized quarks in a transversly polarized
target. This function arises at leading twist in the SSA~\cite{boer}
in conjunction  with the $T$-even unpolarized
fragmentation function. Being $T$-odd, this asymmetry vanishes at
tree level. The important lesson beyond the model calculation, is that,
theoretically, final state interactions are essential for producing
non-zero SSA's. Furthermore, the phenomenological determination of quark
spin distributions can be disentangled from measurements of SSA's.

We have extended this approach to calculate the transversity 
distributions
and corresponding SSA in SIDIS to access {\em transversity}.
Collins~\cite{cnpb92_1} considered one such process, the production 
of pions from transversly
polarized quarks in a transversely polarized target.
The corresponding SSA involves the convolution of
the transversity distribution function and the $T$-odd fragmentation
function, $h_1(x)\star H_1^\perp(z)$~\cite{boer,kotz}. The analogous 
transversity distribution function can be defined through the light-cone 
quark distribution
with gauge link indicated,
\small\be
  s_T^i\Delta f_T(x,k_{\perp})={\frac{1} {2}}\sum_n
\int {\frac{d\xi^- d^2\xi_\perp }
  {(2\pi)^3}} e^{-i(\xi^- k^+-\vec{\xi}_\perp \vec{k}_\perp)} \nonumber 
\\
\langle P|\overline{\psi}(\xi^-,\xi_\perp)|n\rangle 
\langle n|\left(-ie_1\int^\infty_0 A^+(\xi^-,0) d\xi^-
   \right)\gamma^+ \nonumber \\
\gamma^i\gamma^5\psi(0)|P\rangle  + {\rm h. c.},
\nonumber
\ee\normalsize
where $e_1$ is the charge of the struck quark and
$n$ represents intermediate diquark states. In the quark-diquark model
the final state interaction contribution can be evaluated by integrating 
over 
$q^\mu$, the gluon momentum (similar to the calculation in  
refs.~\cite{brodsky,ji,cplb}).  We obtain
\small\be
&&s_{T}^i\Delta
f_T(x,k_{\perp})=\frac{e_1e_2g^2}{2(2\pi)^4}
\frac{1-x}{\Lambda(k_\perp^2)}\quad\quad
\no \\
&&\times\left\{
\left(S^i_T\Big[\bigg(m+xM\bigg)^2+k_\perp^2\bigg]
+2k^i_{\perp }\mathbf{S}_T\cdot\mathbf{k}_\perp\right)\right.
\no \\
&&\times\frac{1}{k^2_\perp+\Lambda(0)^2
+\lambda_g^2}
\left(\ln\frac{\Lambda(k^2_\perp)}{\Lambda(0)}
+\ln\frac{k^2_\perp+\lambda^2_g}{\lambda_g^2}\right)
\no \\
&&-\left.
\left(S^i_T k_\perp^2+2k_\perp^i\mathbf{S}_T\cdot\mathbf{k}_\perp
\right)
\frac{1}{k_\perp^2}\ln\frac{\Lambda(k^2_\perp)}
{\Lambda(0)}
\right\},
\no
\ee\normalsize
\small\be
\Lambda(k_{\perp}^2) = \mathbf{k}_{\perp}^2 +
x(1-x)\left(-M^2+\frac{m^2}{x} + \frac{\lambda^2}{1-x}\right).
\nonumber
\ee\normalsize
The (Abelian) gluon mass (usually chosen at $\lambda_g\approx 1\ GeV$)
is indicative of $\chi SB$ scale and appears here to regulate the IR
divergence.

The first part has the same nucleon spin dependent structure as a tree
level model calculation~\cite{rodriq} - it is leading
twist and a combination of $h_{1T}(x,k_{\perp})$
and $h_{1T}^{\perp}(x,k_{\perp})$. The second
part has a different structure than tree level - it appears as a
rescattering effect only. It is IR finite and, in this model, is
proportional to the one loop result for $f_{1T}^{\perp}$~\cite{ji} 
and ${\cal P}_y$ in BHS. The ratio of $h_{1T}(x,k_{\perp})$
to $h_{1T}^{\perp}(x,k_{\perp})$ will differ from the tree level. 
Integrating
over $k_{\perp}$ leaves $h_1(x)$. This one loop contribution constitutes 
the
next order term in an eikonalization.

When combined with a measure of transversely polarized quarks, the
fragmentation function $H_1^{\perp}(z)$, the
integrated $h_1(x)$ ($h_{1T}(x)$
and the first moment of $h_{1T}^{\perp}(x)$) will contribute to the
observable weighted meson azimuthal asymmetry from a transversely
polarized nucleon~\cite{boer,kotz}.
Weighting by powers of $k_{\perp}$ gives asymmetries in 
$\sin(n\phi_{meson})$.

The $T$-odd structure function $h_1^{\perp}(x,k_{\perp})$ is of more 
interest 
both theoretically, since it vanishes at tree level, and experimentally, 
since its determination does not necessarily involve polarized
nucleons~\cite{boer}. Repeating the calculation above {\em without 
nucleon polarization} leads to the result
\small\be
h_1^\perp(x,k_{\perp})&=&
\frac{e_1e_2g^2}{2(2\pi)^4}
\frac{(m+xM)(1-x)}{\Lambda(k^2_\perp)}
\no
\\
&&\hspace{1cm}\times\, \varepsilon_{+-\perp j}k_{\perp j}
\frac{1}{k_\perp^2}\ln\frac{\Lambda(k^2_\perp)}{\Lambda(0)}.
\no
\ee\normalsize
This is again proportional to the $f_{1T}^{\perp}$ result. It is a 
leading
twist, IR finite result. Being $T$-odd it will appear in SIDIS 
observables
along with $T$-odd fragmentation functions. Many examples of such
observables have been proposed, including SSA's, 
angular distributions of the final hadron and its
polarization~\cite{boer,ggprep,jakob}.


In summary, the Spin-flavor symmetry relates tensor charges to axial 
charges
when supplemented with axial vector dominance. Secondly, axial vector
dominance produces a $\left< k_{\perp}^2 \right>$ factor that appears in
rescattering models in meson photoproduction. Particularly,
transversely polarized nucleon asymmetries in exclusive and inclusive
$\pi$ and $\eta$ photoproduction are interference phenomena that require
rescattering to be non-zero. This is true in SIDIS as well. The exchange
picture for photproduction merges with the struck quark perspective
when rescattering is effective.

The spectator model provides a testing ground for these notions and
yields simple relations. Are they too simple? We are exploring several
related issues. How to specify asymmetries precisely that will focus on
the interesting distribution functions? What are the $k_{\perp}$ 
dependences
from different models? Are there more realistic intermediate 
states (that can carry spin information)?

Acknowledgement: We appreciate communications with S. Brodsky and D.S. 
Hwang.


\begin{thebibliography}{9}
\bibitem{jaffe91}
X.\ Artu and M.\ Mekhfi, Z. Phys. C45 (1990) 669;
R.\ L.\ Jaffe and X.\ Ji, Phys. Rev. Lett. 67 (1991) 552;
Nucl. Phys. B375 (1992) 527.

\bibitem{ralston79}
J. Ralston and D. E. Soper, Nucl. Phys. B152 (1979) 109.

\bibitem{cnpb92_1}
J. C. Collins, Nucl. Phys. B394 (1993) 169.

\bibitem{kotz0} A. M. Kotzinian, Nucl. Phys B441 (1995) 234.

\bibitem{mulders2} R. D. Tangerman and P. J. Mulders,
Phys. Lett. B352 (1995) 129; Phys. Rev. D51 (1995) 3357;
Nucl. Phys. B461 (1996) 197.

\bibitem{hermes} A. Airapetian {\it et al.}, Phys. Rev. Lett.
84 (2000) 4047 (2000); A. Bravar, Nucl. Phys. Proc. Suppl. 79
(1999) 520.

\bibitem{diehl} P. Hoodbhoy and X. Ji, Phys. Rev.D58 (1998), 054006;
M. Diehl, Eur. Phys. J. C19 (2001) 485.

\bibitem{gamb_gold} L. Gamberg and G. R. Goldstein, Phys.
Rev. Lett. 87 (2001) 242001.

\bibitem{gold_owens} G. R. Goldstein and J. F. Owens,
Phys. Rev. D7 (1973) 865; Nucl. Phys. B71 (1974) 461.

\bibitem{brodsky}
S. Brodsky {\it et al.}, Phys. Lett. B530 (2002) 99.

\bibitem{ji} X. Ji and F. Yuan, hep-ph/0206057, 2002; A.V. Belitsky, 
{\it et al.}, hep-ph/0208038.

\bibitem{cplb}
J. C. Collins, Phys. Lett. B536 (2002) 43.

\bibitem{boer} D. Boer and P. J. Mulders, Phys. Rev. D57 (1998) 5780.

\bibitem{kotz} A. M. Kotzinian and P. J. Mulders,
Phys. Lett. B406 (1997) 373.

\bibitem{rodriq} R. Jakob, P.J. Mulders and J. Rodriques, Nucl. Phys.
A626 (1997) 937.

\bibitem{ggprep} L. Gamberg and G. R. Goldstein, in preparation.

\bibitem{jakob} D. Boer, R. Jakob and P.J. Mulders, Nucl. Phys.
564 (2000) 471;
M. Radici, A. Bianconi and R. Jakob, Phys. Rev. D65
(2002) 074037.

\end{thebibliography}
\end{document}